\documentclass{article}

\input pz.sty
\usepackage{longtable}
\usepackage{graphicx}
\usepackage[english]{babel}

\begin{document}

\title{\bf{Variability and Possible Rapid Evolution of the Hot Post-AGB Stars
Hen~3-1347, Hen~3-1428, and LSS~4634}}
\author{V. P. Arkhipova$^{1}$, M.A.Burlak$^{1}$, V.F.Esipov$^{1}$, N. P.
Ikonnikova$^{1}$\thanks{E-mail: ikonnikova@gmail.com}\\
A. Yu. Kniazev$^{1, 2, 3}$, G. V. Komissarova$^{1}$, and A.
Tekola$^{2}$\\
{\small {\it$^{1}$Lomonosov Moscow State University, Sternberg
Astronomical Institute}},\\ {\small {\it Universitetskii pr. 13, Moscow, 119992 Russia}} \\
{\small {\it$^{2}$South African Astronomical Observatory, Cape
Town, South Africa}}\\{\small {\it$^{3}$Southern African Large
Telescope Foundation, Cape Town, South Africa}}}

\date{}

\renewcommand{\abstractname}{ }

\maketitle

\begin{abstract}

We present the results of spectroscopic and photometric
observations for three hot southern-hemisphere post-AGB objects,
Hen~3-1347 = IRAS~17074--1845, Hen~3-1428 = IRAS~17311--4924, and
LSS~4634 = IRAS~18023--3409. In the spectrograms taken with the
1.9-m telescope of the South African Astronomical Observatory
(SAAO) in 2012, we have measured the equivalent widths of the most
prominent spectral lines. Comparison of the new data with those
published previously points to a change in the spectra of
Hen~3-1428 and LSS~4634 in the last 20 years. Based on ASAS data,
we have detected rapid photometric variability in all three stars
with an amplitude up to 0.3 mag, 0.4 mag in the $V$ band. A
similarity between the patterns of variability for the sample
stars and other hot protoplanetary nebulae is pointed out. We
present the results of $UBV$ observations for Hen 3-1347,
according to which the star undergoes rapid irregular brightness
variations with maximum amplitudes $\Delta V$=0.25 mag, $\Delta
B$=0.25 mag, and $\Delta U$=0.30 mag and shows color-magnitude
correlations. Based on archival data, we have traced the
photometric history of the stars over more than 100 years.
Hen~3-1347 and LSS~4634 have exhibited a significant fading on a
long time scale. The revealed brightness and spectrum variations
in the stars, along with evidence for their enhanced mass, may be
indicative of their rapid post-AGB evolution.


\end{abstract}

\newpage

\section*{INTRODUCTION}

The post-asymptotic giant branch (post-AGB) phase of evolution of
intermediate-mass stars ($M_{ZAMS} = 0.8-8 M_{\odot}$) is the
transition period from the asymptotic giant branch (AGB) to
planetary nebula nuclei.

Protoplanetary nebulae are stars of spectral types from late G to
early B with extended atmospheres surrounded by dust envelopes
formed through intense mass loss at previous evolutionary phases.
The dust envelopes gradually dissipate, which is reflected in the
visibility of the stars at optical wavelengths.

The studies of post-AGB stars were begun in the mid-1980s after
the appearance of the IRAS survey. These objects were initially
suspected and subsequently revealed among the supergiant stars
with dust envelopes detected in this survey (Parthasarathy and
Pottasch 1986; van der Veen et al. 1989; Hrivnak et al. 1989).

An important feature of post-AGB stars is their photometric
variability. The type of variability depends on the star's
effective temperature, i.e., on its position on the horizontal
evolutionary track. In particular, F-G supergiants with infrared
excesses exhibit semiregular brightness variations with time
scales from 40 to 130 days caused by their pulsational instability
(Hrivnak and Lu 2000; Kiss et al. 2007; Arkhipova et al. 2010;
Hrivnak et al. 2010). Hotter stars, early B supergiants with
infrared excesses, exhibit photometric variability without any
distinct periodicity with amplitudes from 0$^{m}$.2 to 0$^{m}$.4
in the $V$ band on time scales from one day to several days
(Hrivnak and Lu 2000) and a color-magnitude relation that cannot
be explained by temperature variations. An unstable stellar wind
is considered as the main cause of the photometric variability in
protoplanetary nebulae (Arkhipova et al. 2012). Studies of the
photometric instability and radial velocity curves have revealed
quite a few binary stars among the post-AGB objects (Van Winckel
et al. 2009, 2012).

This paper is devoted to investigating three southern-hemisphere
hot post-AGB stars for which there have been no data on their
photometric behavior until the present time.

{\bf IRAS~17074--1845.} The star was discovered by Henize (1976)
as a result of his survey of southern-hemisphere emission-line
objects and was designated as Hen 3-1347. Subsequently, the object
was identified with the infrared source IRAS 17074--1845 (Dong and
Hu 1991). Based on the similarity between the far-infrared colors
of the star and those of planetary nebulae and taking into account
its high Galactic latitude ($b=+12.^{\circ}26$) and spectral type
Be, Parthasarathy (1993) classified the star as a hot post-AGB
object.

{\bf IRAS~17311--4924.} Henize (1976) discovered this object as an
emission-line star and designated it as Hen 3-1428. Subsequently,
data from the IRAS point source catalogue (Beichmann et al. 1985)
revealed a far-infrared excess in the star. Based on these data,
Parthasarathy and Pottasch (1989) established that the object is
in the post-AGB phase of evolution.

{\bf IRAS 18023--3409.} The star CD-34~12448 was included by
Stephenson and Sanduleak (1971) in the luminous star catalogue as
the object LSS 4634 of spectral type BO+. Based on their analysis
of the data from the IRAS point source catalogue (Beichmann et al.
1985), Preite-Martinez (1989) included IRAS~18023--3409 in the
list of possible new planetary nebulae.

In this paper, we analyzed the ASAS photometric data for all three
stars based on which we detected photometric variability that is
also typical of other hot post-AGB stars. In addition, we carried
out $UBV$ observations for one of the objects, IRAS~17074--1845,
that confirmed the photometric instability of the star.

An important task in investigating post-AGB stars is to reveal as
many rapidly evolving objects as possible to refine their
theoretical evolutionary tracks.

Theoretical calculations of the post-AGB evolution of
intermediate-mass stars predict\\ comparatively short transition
times of the star from an AGB giant to a hot subdwarf and then to
a white dwarf. Depending on the initial mass of the star, the mass
of the remnant star, and the history of mass loss on the AGB, the
HR-diagram crossing time is estimated to be from 100 to several
thousand years (Bl\"{o}cker 1995). Evolutionary changes in a
reasonable time can be noticed only in the most massive post-AGB
objects.

Objects evolving "before our very eyes"\ have already been
discovered among the hot post-AGB stars. These include, for
example, SAO 244567 (Parthasarathy et al. 1993; Arkhipova et al.
2013a) and IRAS 18062+2410 (Arkhipova et al. 1999).

For our study, we selected post-AGB stars that, according to the
estimates by Mello et al. (2012), have masses exceeding the mean
values for planetary nebula nuclei and that, consequently, can
manifest themselves as rapidly evolving objects. For this purpose,
we traced the photometric history of the stars over more than 100
years and analyzed the spectroscopic data over the last 20 years.

Table 1 provides basic information about the objects, namely their
designations, equatorial and Galactic coordinates, $V$ magnitudes,
and spectral types. All three stars have the spectra of B-type
emission-line supergiants, lie at high Galactic latitudes, and are
infrared (IR) sources. The double-humped spectral energy
distribution for these objects is attributable to the radiation
from the stars themselves and the surrounding cold dust envelopes
composed of the material ejected on the AGB. The mass loss from
the stars with a rate of $\dot{M}\sim 10^{-6}-10^{-5} M_{\odot}$
yr$^{-1}$ also continues in the post-AGB phase (Gauba and
Parthasarathy 2004).


\begin{table}

\caption{Program objects}
\bigskip

\begin{tabular}{c|c|c|c|c|c|c|c}
\hline
IRAS&Èìÿ&$\alpha$ (2000.0)&$\delta$ (2000.0)&$l$&$b$&$V$&Sp\\
\hline 17074--1845&BD--18 4436 &
17:10:24&--18:49:01&04.1&+12.3&11.46$^{1}$&B3IIIe$^{3}$, B5Ibe$^{4}$ \\
&Hen 3-1347&&&&&&\\
&LSE 3&&&&&&\\
17311--4924&CD--49 11554&17:35:02&--49:26:26&
341.4&--09.0&10.74$^{2}$&B1IIe$^{5}$,  B3Ie$^{6}$, B1Iae$^{4}$\\
&Hen 3-1428&&&&&&\\
&LSE 76&&&&&&\\
18023--3409&CD--34 12448&18:05:38&--34:09:30&357.6&
--06.3&12.08$^{1}$&B9Ia+e$^{7}$, B2IIIe$^{5}$\\
&LSS 4634&&&&&&\\
\hline

\end{tabular}

$^{1}$ Hog et al. (2000),, $^{2}$ Klare and Neckel (1977), $^{3}$
Gauba et al. (2003), $^{4}$ Carmona et al. (2010),
$^{5}$Parthasarathy et al. (2000), $^{6}$Su\'{a}rez et al. (2006),
$^{7}$Vijapurkar and Drilling (1993).


\end{table}

The objects were included in the Toru\'{n} catalogue of post-AGB
and related objects (Szczerba et al. 2007). IRAS~17074--1845 and
IRAS~17311--4924 are contained in the spectroscopic atlas of
Su\'{a}rez et al. (2006) as transition sources from post- AGB
stars to planetary nebulae. Previously, these objects were
investigated together with other hot post-AGB stars. Gauba et al.
(2003) constructed the spectral energy distribution for them and
estimated the dust temperature, the distances to the stars, the
mass loss rates, and the dynamical ages from the tip of the AGB.
Based on ultraviolet (UV) observations, Gauba and Parthasarathy
(2003) determined the parameters of the stars and estimated their
masses by comparing their data with theoretical evolutionary
tracks from Sch\"{o}nberner (1983, 1987). The properties of the
dust envelope around IRAS~17074--1845 were investigated by
Cerrigone et al. (2009) based on IR photometry and spectroscopy
obtained with the \emph{Spitzer} Space Telescope.

\section*{SPECTROSCOPIC OBSERVATIONS}

Our spectroscopic observations of the program objects were carried
out in May 2012 at the 1.9-m telescope of the South African
Astronomical Observatory (SAAO) with a long-slit spectrograph at
the Cassegrain focus. The slit was $\sim3^{\prime}$ in length and
1.5$^{\prime\prime}$ in width; the scale along the slit was
0.7$^{\prime\prime}$/pixel. The detector was an SITe
266$\times$1798-pixel CCD array. A 300 lines/mm grism was used in
the spectral range $\lambda$ 3500--7200 \AA. The actual spectral
resolution was FWHM = 4.5 \AA. Spectra of a Cu--Ar-filled lamp
were taken to calibrate the wavelengths after each observation.
Bias and flat-field images were also obtained for each night of
observations to perform the standard reduction of two-dimensional
spectra. The log of spectroscopic observations is given in Table
2, where the object names, dates of observation, exposure times,
and airmasses are listed. Figure 1 shows the spectra of the
program objects normalized to the continuum in the range $\lambda$
3750--7500 \AA.

\begin{figure}
\centering
\includegraphics[scale=1.1]{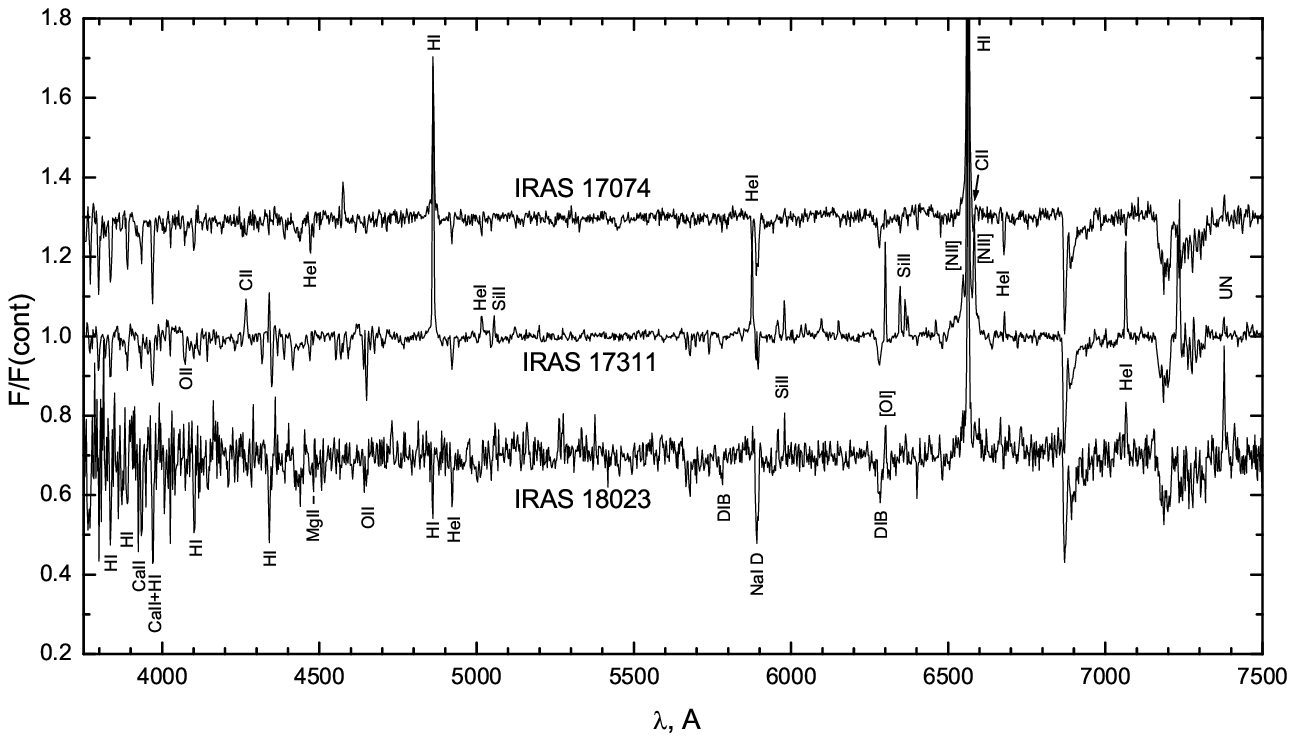}
\caption{Spectra of the program stars normalized to the continuum.
The level of normalized spectra for IRAS 17074--1845 and IRAS
18023--3409 is arbitrarily shifted.} \label{fig1}
\end{figure}

\begin{table}
\begin{center}
\caption{Log of observations}\label{log}
\begin{tabular}{llll} \hline \rule[-9pt]{0pt}{24pt}
IRAS  & Date & Exp, s & $F(z)$\\
\hline \rule{0pt}{14pt}\hspace{-1.2mm} 17074--1845 & 2012-05-12& 600 & 1.04 \\
 & 2012-05-22 & 900 & 1.15 \\
17311--4924 & 2012-05-12& 2$\times$600 & 1.05\\
18023--3409 & 2012-05-11& 600 & 1.05\\

\hline
\end{tabular}
\end{center}
\end{table}

One of the objects, IRAS~17074--1845, was also observed with a
125-cm reflector at the Crimean Station of the Sternberg
Astronomical Institute (Moscow State University) on May 22, 2012,
using a spectrograph with a 600 lines mm$^{-1}$ diffraction
grating and a 4$^{\prime\prime}$ slit. The detector was an ST-402
CCD array (the CCD size is 765$\times$510 pixels, the pixel size
is 9$\times$9 $\mu$m). The spectral resolution (FWHM) was 7.4 \AA\
in the range $\lambda$ 4000--7200 \AA. The spectra were processed
using the standard CCDOPS code and the SPE code developed by S.G.
Sergeev at the Crimean Astrophysical Observatory.

In addition, we used the spectroscopic data for IRAS~17074--1845
and IRAS~17311--4924 from the appendix to the "Spectroscopic atlas
of post- AGB stars and planetary nebulae"\ by Su\'{a}rez et al.
(2006) provided in the VizieR electronic database (J/A+A/458/173).
For IRAS~17074--1845, there is a record of the spectrum taken in
the period March 13.17, 1994, in the spectral range $\lambda$
3285--10~980 \AA\ and IRAS~17311--4924 was observed on March
10.13, 1993, in the range $\lambda$ 3321--11~015 \AA. The
observations were carried out in Chile with a 1.5-m telescope at
the La Silla Observatory of the European Southern Observatory
(ESO) using the Boller-Chivens spectrograph. The formal resolution
was 3.74 \AA\ per pixel, which is close to our formal resolution
($\sim$ 2.3 \AA\ per pixel for the spectra taken at SAAO and in
Crimea). The spectra are given in absolute units (erg cm$^{-2}$
s$^{-1}$ \AA$^{-1}$).

Based on the spectra that we took at SAAO and in Crimea and on the
spectra from the atlas by Su\'{a}rez et al. (2006), we measured
the line equivalent widths. For this purpose, the profile of each
line was fitted by a Gaussian after the spectrum normalization.
The equivalent width was assumed to be equal to the area of the
derived Gaussian. The total error in the line equivalent width was
calculated as a quadratic sum of the fitting error and the
continuum placement error. The latter was obtained using the AMD
algorithm (Kniazev et al. 2004), which gives an estimate of the
continuum noise per one point of the spectrum. The total error of
the continuum for each line was found in a wavelength interval
equal to 2$\times$line FWHM.

{\bf Hen~3-1347 = IRAS~17074--1845.} In the low-resolution
spectrograms from the appendix to the paper by Su\'{a}rez et al.
(2006) and those taken at SAAO and in Crimea, the emission
spectrum of IRAS~17074--1845 is represented by the Balmer
H$\alpha$ and H$\beta$ lines and very weak [O I] lines. There are
no forbidden [NII] lines. The H$\gamma$, H$\delta$, HeI, CII
($\lambda$ 4267, 5045, 6578+6582, 7231+7236 \AA ), Si II
($\lambda$ 6347 and 6371 \AA ), and Ca II H and K lines are
observed in absorption. The spectrum also exhibits individual OII
absorption lines. We measured the equivalent widths of the most
prominent emission and absorption lines in the spectrograms of
IRAS~17074--1845 and give them in Table 3. In the table, the
equivalent widths of the emission and absorption lines are
positive and negative, respectively.

As can be seen from Table 3, the line equivalent widths in the
spectrograms taken at different dates agree well between
themselves. Thus, the stars's spectrum underwent no significant
changes from 1994 to 2012.

\begin{table}
\begin{center}
\caption{Line equivalent widths in the spectrum of
IRAS~17074--1845.} {\small
\begin{tabular}{ccccc} \hline $\lambda$, \AA & Ion
&\multicolumn{3}{c} {$W\pm \sigma_W$, \AA} \\
&&2012-05-11 & 2012-05-22 & 1994-03\\ \hline
4471 & HeI & -0.54$\pm$0.10 &  & \\
4481 & MgII & -0.25$\pm$0.08 &  & \\
4713 & HeI & -0.17$\pm$0.08 &  & \\
4861 & H$\beta$ & 2.15$\pm$0.11 & 2.36$\pm$0.21 & 2.22$\pm$0.19 \\
4921 & HeI & -0.4$\pm$0.08 &  -0.41$\pm$0.11 & \\
5016 & HeI & -0.25$\pm$0.08 &  & \\
5876 & HeI & -0.4$\pm$0.08 & -0.16$\pm$0.09 & \\
5893 & NaI & -1.55$\pm$0.11 & -1.55$\pm$0.16 & -1.1$\pm$1.0\\
6563 & H$\alpha$ & 15.9$\pm$0.3 & 16$\pm$0.7 & 17.8$\pm$0.7 \\
6678 & HeI & -0.53$\pm$0.10 & -0.4$\pm$0.11 & -0.37$\pm$0.06 \\
7065 & HeI & -0.24$\pm$0.07 &  & -0.13$\pm$0.06 \\
7378 & UN & 0.29$\pm$0.07 & & 0.25$\pm$0.09 \\
 \hline
\end{tabular}
}
\end{center}
\end{table}

Let us also present the previously published results of
spectroscopic observations for\\ IRAS~17074--1845.

Parthasarathy et al. (2000) took a spectrum of the star in April
1994 using a spectrograph and a two-dimensional photon-counting
detector in the range $\lambda$ 3800--5000 \AA\ with a resolution
of about 3.5 \AA. These authors estimated the spectral type of
IRAS~17074--1845 as B3IIe and reported that the H$\beta$ and
H$\gamma$ were observed in emission. Based on the spectrum taken
also in 1994, Su\'{a}rez et al. (2006) determined the spectral
type of the star as B3Ie.

On August 4, 2007, a spectrum of IRAS~17074--1845 was taken with a
high resolution ($R\sim$45~000) (Carmona et al. 2010). These
authors reported that the H$\alpha$, H$\beta$, H$\gamma$, and
H$\delta$ were observed in emission, although H$\alpha$ in the
table to the paper is designated as a line with a P~Cyg profile
for which the equivalent widths of the emission ($W_{\lambda}$ =
12 \AA ) and absorption ($W_{\lambda}$=--0.8 \AA ) components were
measured. The He~I $\lambda$ 4009, 4121, 4713, 4922, 5016, 5876,
and 6678 \AA\ lines are observed in absorption; there are no He II
lines. The spectrum also features the N~II $\lambda$ 3995 \AA,
Si~II $\lambda$ 4128 and 4131 \AA, C~II $\lambda$ 4267 \AA, Si~III
$\lambda$ 4553, 4568, 4575 \AA\ absorption lines. The Mg~II
$\lambda$ 4481 \AA\ absorption line is slightly stronger than He~I
$\lambda$ 4471 \AA. By applying various criteria for spectral
classification, the authors estimated the spectral type of the
star as B5Ibe.

Mello et al. (2012) took a spectrum of IRAS~17074--1845 with a
resolution $R\sim$48 000 on May 22, 2008, in the range $\lambda$
3600--9200 \AA. In this spectrum, the H$\gamma$, H$\beta$, and
H$\alpha$ lines have P~Cyg profiles with a broad blueshifted
absorption. Some of the He~I lines exhibit emission components.
There are forbidden Fe~II lines. The authors modeled the spectrum
and obtained the stellar parameters ($T_{eff} = 15~300\pm$600 K,
$\log g$ = 2.05, $\xi$ = 10 km s$^{-1}$).

Thus, the spectrum of IRAS~17074--1845 was observed in 1994, 2007,
2008, and 2012. Its spectra were taken at different telescopes and
with different resolutions. High-resolution observations revealed
P~Cyg profiles for individual lines in the star's spectrum,
suggesting the presence of a stellar wind. The equivalent widths
of the H$\alpha$ and H$\beta$ lines measured in low-resolution
spectra underwent no changes exceeding the observational errors.
The differences in the estimates of the spectral type for the star
from B3 to B5 can reflect both the real changes in the spectrum
and the difference between the applied criteria for spectral
classification.

{\bf Hen~3-1428 = IRAS~17311--4924}. The spectroscopic
observations of IRAS~17311--4924 have been carried out repeatedly.

Having analyzed the spectrum taken in April 1994 in the range
$\lambda$ 3800--5000 \AA\ with a resolution of about 3.5 \AA\ ,
Parthasarathy et al. (2000) assigned the spectral type B1IIe to
IRAS~17311--4924.

Based on the spectrum taken in the period March 10-13, 1993,
Su\'{a}rez et al. (2006) determined the spectral type of the star
as B3Ie.

The high-resolution spectroscopy in the range $\lambda$ 4900--8250
\AA\ performed by Sarkar et al. (2005) on June 22, 2002, allowed
one to identify numerous emission and absorption lines in the
spectrum of IRAS~17311--4924, to measure their equivalent widths,
to reveal P~Cyg profiles for the H$\alpha$, He~I, Fe~III, and C~II
lines, and to estimate the envelope expansion velocity, the
object's heliocentric velocity, and the stellar wind velocity.

Carmona et al. (2012) observed IRAS~17311--4924 on August 7, 2007,
with a high-resolution ($R$ = 45~000) spectrograph in the range
$\lambda$ 3500--9200 \AA. The Balmer and Paschen lines in their
spectrum are observed in emission with P Cyg profiles. There are
no He~II lines. The He~I $\lambda$ 4471, 4713, 5016, 5876 \AA\
lines also have P Cyg profiles. Forbidden lines are present in the
star's spectrum: [O~I] $\lambda$ 5577, 6300, 6364 \AA; [N~II]
$\lambda$ 6548 and 6584 \AA. The C~II $\lambda$ 6578 \AA\ line
with a P~Cyg profile is seen in the figure given in the paper; the
authors took the second doublet line, C~II $\lambda$ 6583 \AA, for
the P~Cyg profile of [N~II] $\lambda$ 6584 \AA. Analysis of the
spectrum allowed the authors to determine the spectral type as
B1Iae, to which $T_{eff}$=20~800 K corresponds (Lang et al. 1991).

Mello et al. (2012) took a spectrum of IRAS~17311--4924 with a
resolution $R\sim$ 48 000 on May 23, 2008, in the range $\lambda$
3600-9200 \AA. These authors pointed out significant changes
compared to the data from Sarkar et al. (2005): the intensity of
the H$\alpha$ emission line increased considerably; the emission
components of the He~I lines as well as the Fe~III and vanadium
V~I emission lines associated with the circumstellar material
changed. The parameters of the star derived by Mello et al.
(2012), $T_{eff}$=20~500 $\pm$ 500 K and $\log g$=2.35, agree well
with the estimate of its spectral type B1Iae (Carmona et al.
2012).

In the spectrogram that we took on May 5, 2012, the H$\alpha$,
H$\beta$, and H$\gamma$ lines are emission ones and H$\delta$ is
observed in absorption. Since the spectral resolution is
insufficient, we cannot reveal any lines with P~Cyg profiles. The
spectrum exhibits both emission ($\lambda$ 5016, 5876, 6678, 7065,
7281 \AA) and absorption ($\lambda$ 3820, 3927, 4028, 4120, 4144,
4388, 4471, 4922 \AA) He~I lines. Fairly strong O~II absorption
lines are observed; there are several N~II and N~III absorption
lines. The C~II~$\lambda$ 4267, 5122, 6095+6099 \AA\ lines and the
very strong $\lambda$ 7231+7236 \AA\ doublet are observed in
emission; the C~III~($\lambda$~4647, 4650, 4651 \AA) lines are
observed in absorption. The situation with silicon is similar: the
Si~II~($\lambda$~5056, 5958, 5979, 6347, 6371 \AA) lines are
mostly emission ones, while the Si~III lines are absorption ones.
The Ca~II H and K absorption lines are clearly seen. Out of the
forbidden lines, only [O~I]~$\lambda$~6300, 6364 \AA\ and
[N~II]~$\lambda$~6548, 6584 \AA\ are seen. The auroral [O~I]
$\lambda$ 5577 \AA\ line is not detected in our spectrum. We
estimated the spectral type from the ratio of the He~I and Mg~II
lines ($\lambda$~4471 and 4481 \AA) to be B1.5-2.

We measured the equivalent widths of the strongest emission and
absorption lines in the spectrum of IRAS~17311--4924 and provide
them in Table~4, along with the equivalent widths of the
individual lines measured in the spectrum from the appendix to the
paper by Su\'{a}rez et al. (2006). It can be seen from the table
that the equivalent widths of the He~I lines had changed only
slightly since 1993, while the equivalent widths of H$\beta$ and
H$\alpha$ and the forbidden [O~I] $\lambda$ 6300 \AA\ and [N~II]
$\lambda$ 6584 \AA\ lines had increased noticeably by 2012.


\begin{table}
\begin{center}
\caption{Line equivalent widths in the spectrum of
IRAS~17311--4924} {\small
\begin{tabular}{cccc|cccc} \hline
$\lambda$, \AA & Ion & \multicolumn{2}{c|}{$W\pm\sigma_W$, \AA} &
$\lambda$, \AA & Ion & \multicolumn{2}{c}{$W\pm\sigma_W$, \AA} \\
&&2012&1993&&&2012&1993\\
 \hline
3798 & H10 & -0.53$\pm$0.10 &  & 5016 & HeI & 0.39$\pm$0.07 & \\
3820 & HeI & -0.32$\pm$0.05 &  & 5048 & CII & -0.18$\pm$0.06 & \\
3835 & H9 & -0.72$\pm$0.06 &  & 5056 & SiII & 0.26$\pm$0.06 & \\
3889 & H8 & -0.47$\pm$0.12 &  & 5122 & CII & 0.18$\pm$0.06 & \\
3927 & HeI, OII & -0.17$\pm$0.05 &  & 5666 & NII & -0.15$\pm$0.05 &\\
3934 & CaII & -0.46$\pm$0.06 &  & 5680 & NII & -0.35$\pm$0.07 & \\
3956 & SiII & -0.34$\pm$0.08 &  & 5740 & SiIII & -0.24$\pm$0.07 & \\
3969 & CaII, H$\epsilon$ & -1.14$\pm$0.07 &  & 5781 & DIB & -0.25$\pm$0.07 & \\
4072 & OII & -0.59$\pm$0.08 &  & 5876 & HeI & 1.4$\pm$0.12 & 1.54$\pm$0.12 \\
4102 & H$\delta$ & -0.49$\pm$0.10 &  & 5890 & NaI & -0.27$\pm$0.06 & \\
4120 & HeI & -0.42$\pm$0.09 &  & 5896 & NaI & -0.38$\pm$0.06 &  \\
4144 & HeI & -0.27$\pm$0.06 &  & 5958 & SiII & 0.32$\pm$0.07 & 0.35$\pm$0.08 \\
4267 & CII & 0.67$\pm$0.06 &  & 5979 & SiII & 0.41$\pm$0.05 & 0.33$\pm$0.10 \\
4317 & OII & -0.38$\pm$0.06 &  & 6033 & FeIII & 0.15$\pm$0.05 & \\
4340 & H$\gamma$ & 0.48$\pm$0.09 &  & 6047 & OI & 0.18$\pm$0.05 &\\
4349 & OII & -0.86$\pm$0.11 &  & 6095 & CII & 0.51$\pm$0.09 & \\
4367 & OII & -0.18$\pm$0.05 &  & 6151 & UN & 0.22$\pm$0.05 & \\
4388 & HeI & -0.31$\pm$0.06 &  & 6300 & [OI] & 1.1$\pm$0.07 & 0.72$\pm$0.14 \\
4415 & OII & -0.69$\pm$0.07 &  & 6347 & SiII & 0.67$\pm$0.06 & \\
4470 & HeI & -0.4$\pm$0.08 &  & 6364 & [OI] & 0.44$\pm$0.04 & \\
4552 & SiIII & -0.33$\pm$0.05 &  & 6371 & SiII & 0.29$\pm$0.05 &\\
4568 & SiIII & -0.31$\pm$0.06 &  & 6548 & [NII] & 0.61$\pm$0.21 & \\
4575 & SiIII & -0.165$\pm$0.05 &  & 6563 & H$\alpha$ & 23$\pm$0.2 & 17.2$\pm$0.7 \\
4591 & OII & -0.42$\pm$0.09 &  & 6584 & [NII] & 1.4$\pm$0.21 & 0.53$\pm$0.5\\
4639 & OII & -0.43$\pm$0.07 &  & 6460 & UN & 0.2$\pm$0.05 & \\
4650 & OII & -0.87$\pm$0.10 &  & 6678 & HeI & 0.28$\pm$0.05 & 0.31$\pm$0.09\\
4661 & OII & -0.13$\pm$0.07 &  & 6640 & OII & -0.28$\pm$0.07 &\\
4676 & OII & -0.18$\pm$0.06 &  & 6722 & OII & -0.15$\pm$0.05 &\\
4861 & H$\beta$ & 3.81$\pm$0.11 & 2.74$\pm$0.18 & 7065 & HeI & 1.31$\pm$0.07 & 1.31$\pm$0.15\\
4891 & OII & -0.19$\pm$0.06 &  & 7231 & CII & 1.26$\pm$0.17 & \} 1.6$\pm$0.3\\
4907 & OII & -0.15$\pm$0.06 &  & 7236 & CII & 1.48$\pm$0.16 &\\
4922 & HeI & -0.58$\pm$0.06 &  & 7378 & UN & 0.27$\pm$0.03 &\\

\hline
\end{tabular}
}
\end{center}
\end{table}

Our 2012 data differ significantly from the results of the 2002
observations by Sarkar et al. (2005). The equivalent widths of
both emission and absorption lines that we measured are higher
than those from Sarkar et al. (2005) by a factor of $\sim$1.5-2.
These differences may be associated to a greater extent with the
difference in spectral resolution. However, Mello et al. (2012)
also draw attention to the fact that the H$\alpha$ intensity in
their spectrum increased considerably compared to the data from
Sarkar et al. (2005).

Thus, the spectroscopy for IRAS~17311--4924 obtained from 1993 to
2012 showed, first, variable P~Cyg profiles, suggesting a variable
stellar wind, and, second, a change in the equivalent widths of
H$\beta$ and H$\alpha$ and the forbidden [O~I] and [N~II] lines
with time.

{\bf LSS~4634 = IRAS~1802--3409.} IRAS~18023--3409 is the weakest
and least studied star from our sample. In our spectrum taken on
May 11, 2012, the signal-to-noise ratio is rather low
(S/N$\sim$25); there are very few prominent features in the
spectrum. The H$\alpha$, [O~I], Si~II, He~I $\lambda$ 7065 \AA\
lines and the unidentified $\lambda$ 7378 \AA\ line, which is also
present in the spectra of IRAS~17074--1845 and IRAS~17311--4924
are observed in emission. Among the absorptions, there are the
Ca~II H and K, H$\delta$, H$\gamma$, H$\beta$, He~I $\lambda$ 4921
and 6678 \AA\ as well as the interstellar and circumstellar Na~I D
line. The line equivalent widths that we were able to measure are
given in Table 5.


\begin{table}
\begin{center}
\caption{Line equivalent widths in the spectrum of
IRAS~18023--3409.} {\small
\begin{tabular}{ccp{2cm}} \hline $\lambda$, \AA & Ion
& $W\pm\sigma_W$, \AA \\
\hline
4102 & H$\delta$ & -1.18$\pm$0.29\\
4340 & H$\gamma$ & -1.24$\pm$0.17\\
4861 & H$\beta$ & -0.72$\pm$0.15\\
4922 & HeI & -0.78$\pm$0.18\\
5893 & NaI & -2.12$\pm$0.55\\
6563 & H$\alpha$ & 8.93$\pm$0.44\\
7065 & HeI & 0.75$\pm$0.16\\
7378 & UN & 1.44$\pm$0.22\\
\hline
\end{tabular}
}
\end{center}
\end{table}


Parthasarathy et al. (2000) took a spectrum of IRAS~18023--3409 in
April 1994. The authors pointed out that H$\beta$ was observed in
emission and that the H$\gamma$ absorption line was filled with
emission. The spectral type was estimated to be B2IIIe.
Previously, Vijapurkar and Drilling (1993) assigned a later
spectral type to the star, B9Ia+e$_{+2}$, based on the spectrum
taken in 1978.

Mello et al. (2012) carried out high-resolution spectroscopic
observations on May 20, 2008. The spectrum of IRAS~18023--3409
exhibits strong forbidden [Fe~II] lines, weak nebular [N~II]
lines; the H$\gamma$ and H$\beta$ lines have P~Cyg profiles. A
nitrogen overabundance correlating with an enhanced helium
abundance was detected. The model parameters of the star given in
Mello et al. (2012), $T_{eff}$=19~400~K and $\log g$=2.28,
correspond to the spectral type B2-3I in the calibration by
Strai\v{z}ys (1982).

\section*{PHOTOMETRIC OBSERVATIONS}
\subsection*{ASAS-3 Observations}

All stars from our program fell within the field of view of the
All Sky Automated Survey (ASAS) (Pojmanski 2002;
\texttt{http://www.astrouw.edu.pl/asas/}). The observations in the
ASAS-3 system have been carried out for more than 10 years at the
Las Campanas (Chile) telescopes in an automatic mode in a
photometric $V$ band close to Johnson's standard $V$.

The sample stars were observed in 2001-2009. Table~6 provides data
on the ASAS observations, namely the mean $<V>$, the mean error of
a single measurement, the maximum amplitude of brightness
variations, and the time interval.


\begin{table}
\centering

\caption{Log of ASAS observations}
\bigskip

\begin{tabular}{c|c|c|c|c}
\hline
IRAS&$<V>$&$\sigma_{V}$&$\Delta V$&JD 2400000+\\
\hline
17074--1845&11.67&0.04&0.4&51936$\div$54883\\
17311--4924&10.84&0.04&0.3&51933$\div$55104\\
18023--3409&12.41&0.05&0.4&51948$\div$55105\\
\hline

\end{tabular}

\end{table}


To analyze the data, we used the measurements made with aperture 1
(15$^{\prime\prime}$) and marked in the ASAS database by symbol A
(good quality).

Figure 2 shows the light curves derived from ASAS data. The
pattern of variability for all three stars is similar in both
characteristic time scales of brightness variations and
oscillation amplitudes. The stars exhibit rapid, from night to
night, photometric variability with amplitudes up to
0$^{m}$.3-0$^{m}$.4. Our search for a periodicity did not lead to
a positive result. Apart from rapid chaotic instability,
IRAS~17311--4924 and IRAS~18023--3409 showed a change in the
yearly mean brightness.

\begin{figure}
\centering
\includegraphics[scale=0.8]{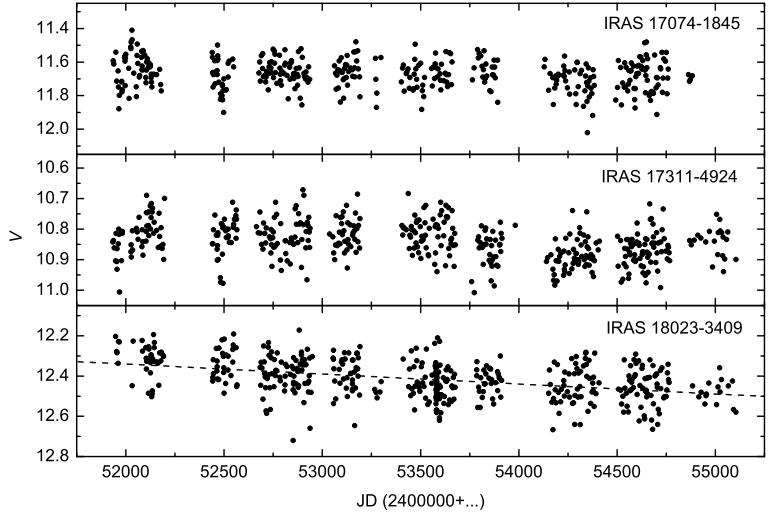}
\caption{Light curves derived from ASAS data. For
IRAS~18023--3409, the dashed line indicates a linear fit to the
data.} \label{fig2}
\end{figure}

\subsection*{$UBV$ Observations of IRAS~17074--1845}

IRAS~17074--1845, the northernmost star in our program, is
accessible to observations at the Crimean Station of the Sternberg
Astronomical Institute (Moscow State University). During
2012-2013, we obtained 37 magnitude estimates for the star in the
$UBV$ bands. The measurements were made with a 60-cm Zeiss-1
telescope using a photoelectric $UBV$ photometer designed by Lyuty
(1971). The photometric system of the photometer is close to
Johnson's one. We used the star HD~155648, whose $UBV$ magnitudes
($V=8^{m}.57$, $B=8^{m}.91$, and $U=9^{m}.06$) were taken from
Wehinger and Hidajat (1973), as a photometric standard. As a rule,
the measurement errors did not exceed $\sigma=0^{m}.01$ in all
three bands. Table~7 gives the $UBV$ observations of
IRAS~17074--1845 and Fig.~3 shows the light curves of the star.
The mean magnitudes and colors of IRAS~17074--1845 in 2011-2012
were $V=11^{m}.65\pm0.06$, $B=11^{m}.93\pm0.06$,
$U=11^{m}.34\pm0.07$, $B-V=0.28\pm0.02$, and $U-B=-0.59\pm0.03$.
The $V$ magnitude virtually coincided with the star's mean
magnitude from the ASAS data in 2001-2009, while the amplitude of
brightness variations from our $UBV$ photometry is smaller, which
is probably because the error of our measurements is smaller than
that of the ASAS observations.

\begin{figure}
\centering
\includegraphics[scale=1.0]{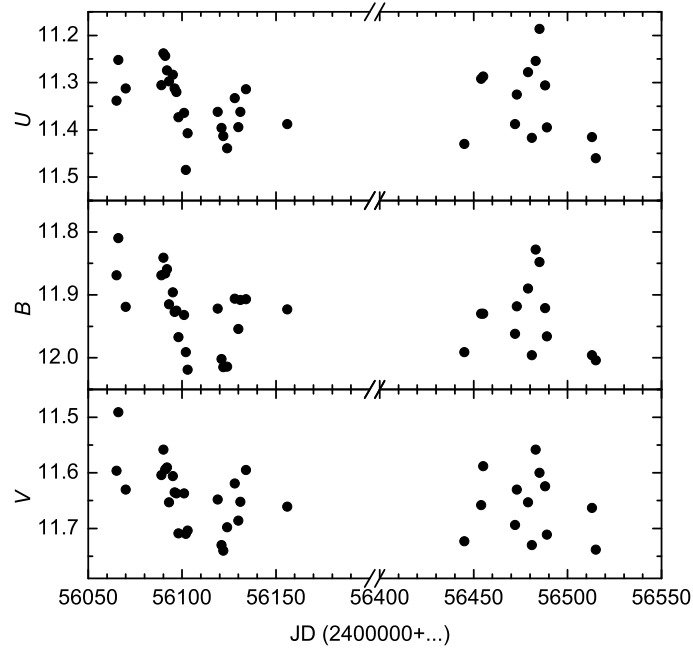}
\caption{$UBV$ light curves of IRAS~17074--1845.} \label{fig3}
\end{figure}


\begin{center}
\begin{longtable}{|c|c|c|c|}
\caption{$UBV$ observations of IRAS~17074--1845 in
2012-2013}\\
\hline
JD& $V$ & $B-V$ & $U-B$\\
\hline
\endfirsthead

\multicolumn{4}{c}{\tablename\ \thetable{}}\\
\hline
JD& $V$ & $B-V$ & $U-B$\\
\hline
\endhead

\hline
\endfoot

\hline
\endlastfoot

\hline

2456065&   11.596&  0.273&   -0.531\\
2456066&   11.491&  0.319&   -0.558\\
2456070&   11.630&  0.289&   -0.607\\
2456089&   11.604&  0.265&   -0.564\\
2456090&   11.558&  0.283&   -0.603\\
2456091&   11.594&  0.272&   -0.623\\
2456092&   11.590&  0.269&   -0.585\\
2456093&   11.653&  0.262&   -0.618\\
2456095&   11.606&  0.290&   -0.613\\
2456096&   11.635&  0.292&   -0.615\\
2456097&   11.637&  0.288&   -0.605\\
2456098&   11.709&  0.258&   -0.594\\
2456101&   11.637&  0.295&   -0.568\\
2456102&   11.710&  0.281&   -0.506\\
2456103&   11.704&  0.315&   -0.612\\
2456119&   11.648&  0.274&   -0.560\\
2456121&   11.730&  0.272&   -0.606\\
2456122&   11.740&  0.275&   -0.602\\
2456124&   11.698&  0.316&   -0.575\\
2456128&   11.619&  0.287&   -0.573\\
2456130&   11.686&  0.268&   -0.560\\
2456131&   11.652&  0.256&   -0.546\\
2456134&   11.595&  0.312&   -0.593\\
2456156&   11.661&  0.262&   -0.535\\
2456445&   11.723&  0.268&   -0.561\\
2456454&   11.658&  0.272&   -0.638\\
2456455&   11.588&  0.342&   -0.643\\
2456472&   11.694&  0.268&   -0.574\\
2456473&   11.630&  0.288&   -0.593\\
2456479&   11.653&  0.237&   -0.612\\
2456481&   11.730&  0.266&   -0.579\\
2456483&   11.558&  0.270&   -0.574\\
2456485&   11.600&  0.248&   -0.662\\
2456488&   11.624&  0.297&   -0.615\\
2456489&   11.711&  0.255&   -0.571\\
2456513&   11.663&  0.333&   -0.581\\
2456515&   11.738&  0.266&   -0.544\\

\hline

\end{longtable}
\end{center}


Our $UBV$ photometry points to rapid photometric variability of
the star from night to night with maximum amplitudes $\Delta
V=0.25$, $\Delta B=0.25$, and $\Delta U=0.3$.

The photometric instability is accompanied by a change in colors:
as the star brightens, $B-V$ may increase, while $U-B$ definitely
decreases. This is illustrated by the color-magnitude diagrams
(Fig.~4). We found such correlations for other hot post-AGB stars
(see, e.g., Arkhipova et al. 2012).

\begin{figure}
\centering
\includegraphics[scale=0.8]{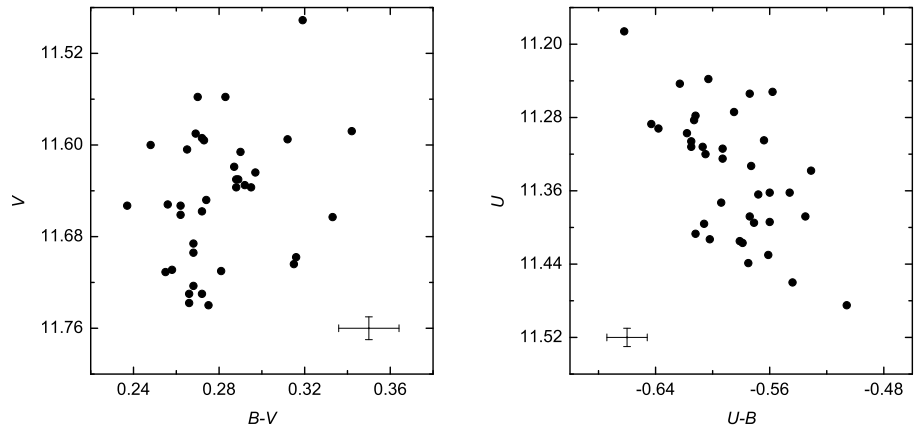}
\caption{Color-magnitude diagrams derived from the $UBV$ data for
IRAS~17074--1845.} \label{fig4}
\end{figure}

We can estimate the color excess from our $UBV$ data. Taking the
spectral type B3I and the normal colors $(B-V)_0=-0.13$ and
$(U-B)_{0}=-0.85$ in the calibration of Strai\v{z}ys (1982) for
the star, we obtained an estimate of $E(B-V)=0.41\pm0.02$.
Comparison of this value with the maximum value of $E(B-V)=0.28\pm
0.01$ (\texttt{http://irsa.ipac.caltech.edu/applications/DUST/})
toward IRAS~17074--1845 derived from the maps by Schlegel (1998)
leads us to conclude that there exists some fraction,
$\sim0^{m}.01$, of circumstellar extinction. Given the ambiguity
in determining the star's spectral type by different authors (see
Table~1) and the variability of the $B-V$ color, the error of the
derived circumstellar extinction can be significant.

\subsection*{Photometric History of the Program Stars}

All three stars from our sample enter into historical catalogues:
BD--18$^{\circ}$4436=IRAS~17074--1845 is contained in the Bonner
Durchmusterung (BD), while CD--49$^{\circ}$11554=IRAS~17311--4924
and CD--34$^{\circ}12448$=IRAS~18023--3409 are contained in the
Cordoba Durchmusterung (CD), which allowed the photometric history
of the stars to be traced over more than 100 years. However, the
visual magnitudes from the BD and CD catalogues should be analyzed
before they are compared with the present-day photometric data.

\begin{figure}
\includegraphics[scale=0.9]{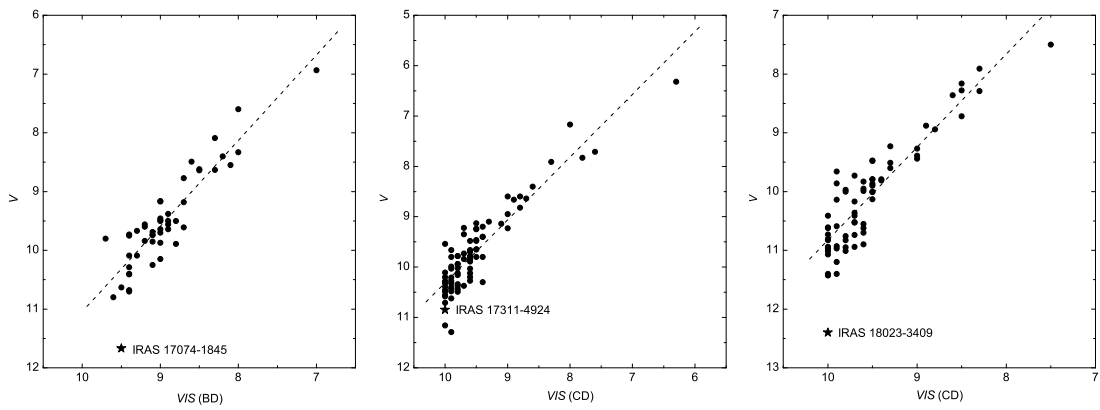}
\caption{Relationship between the $V$ magnitudes from the Tycho-2
catalogue (Hog et al. 2000) and the visual magnitudes from the CD
and BD catalogues for the program objects (asterisks) and stars
from their neighborhoods (dots). The dashed lines indicate linear
fits to the data for the stars from the neighborhoods.}
\label{fig5}
\end{figure}

We selected stars contained in the CD or BD catalogues from a
$1^{\circ}\times1^{\circ}$ neighborhood of each program object.
Figure~5 shows the relationship between the $V$ magnitudes
transformed from the $V_{T}$ magnitudes from the Tycho-2 catalogue
(Hog et al. 2000) and the visual magnitudes ($VIS$) from the CD
and BD catalogues of the program objects and stars from their
neighborhoods. The relationships between $V$ and $VIS$ of the
stars from the neighborhoods obey linear laws, according to which
$VIS$ of each program object are transformed into the following
magnitudes:

IRAS~17074--1845 $VIS=9^{m}.5$ $\mapsto$  $V=10^{m}.3\pm0.8$;

IRAS~17311--4924 $VIS=10^{m}.0$ $\mapsto$ $V=10^{m}.3\pm0.6$;

IRAS~18023--3409 $VIS=10^{m}.0$ $\mapsto$ $V=10^{m}.8\pm0.8$.

{\bf IRAS~17074--1845.} At the epoch of the BD catalogue (the
1950-1960s), the star BD--18$^{\circ}$4436= IRAS~17074--1845 had a
brightness corresponding to $V=10^{m}.3\pm0.8$.

Later observations give the following magnitude estimates for the
star:

$V=11^{m}.46\pm0.15$ and $B=11^{m}.96\pm0.16$ (1989-1992, Tycho-2;
Hog et al. 2000);

$V=11^{m}.55\pm0.05$ (April 10, 2000; Gauba et al. 2003);

$<V>=11^{m}.67\pm0.10$ (2001-2009, ASAS);

$<V>=11^{m}.65\pm0.06$, $<B>=11^{m}.93\pm0.07$,
$<U>=11^{m}.34\pm0.08$ (2012-2013, our data).

Thus, at the present epoch the star has become considerably
fainter than at the epoch of the BD catalogue.

The difference between the star's near-IR magnitudes from
different data is also noteworthy.

The first observations of the star in the $JHK$ bands refer to the
time interval May 23, 1989-June 5, 1989, when Garc\'{i}a-Lario et
al. (1997) obtained the following magnitudes for IRAS~17074--1845:

$J=11.33\pm0.13$, $H=12.06\pm0.16$, $K=11.02\pm0.11$.

Nine years later (on May 1, 1998), according to the 2MASS data
(Skrutskie et al. 2006), the star became considerably brighter:

$J=10.583\pm0.023$, $H=10.456\pm0.024$, $K_{s}=10.357\pm0.026$.

According to the data from Garc\'{i}a-Lario et al. (1997), the
source IRAS~17074--1845 has anomalous colors, $J-H=-0.73$ and
$H-K=1.04$, that are typical of neither post-AGB stars nor
planetary nebulae. However, with the colors $J-H=0.157$ and
$H-K=0.099$ from 2MASS, IRAS~17074--1845 on the $(J-H)-(H-K)$
color-color diagram falls into the region of hot post-AGB stars
with an insignificant IR excess belonging to the dust envelope.

For another star from our sample, IRAS~17311--4924, the IR
photometry from Garc\'{i}a-Lario et al. (1997) and 2MASS showed
complete agreement, rejecting the assumption about a significant
difference between the two photometric systems.

The DENIS infrared observations of IRAS~17074--1845 (Epstein et
al. 1999) performed on July 7, 1997, $J=10.907\pm0.060$ and
$K_{s}=10.634\pm0.080$, show a lower brightness of the star than
the 2MASS data. The difference exceeds 3$\sigma$.

Based on the above $JHK$ photometry, we can assume the star to be
variable in the near infrared.

{\bf IRAS~17311--4924.}The brightest star from our sample,
CD--49$^{\circ}$11554=IRAS~17311--4924, at the epoch of the CD
catalogue had a visual magnitude $VIS=10^{m}.0$ that corresponds
to $V=10^{m}.3\pm0.6$ and differs insignificantly from the
subsequent magnitude estimates.

The first $UBV$ observations of the star belong to Klare and
Neckel (1977) and refer to 1973-1974:

$U=10^{m}.57$, $B=11^{m}$.15, $V=10^{m}$.74.

Subsequently, in 1979 and 1980, Kozok (1985) obtained three
magnitude estimates for IRAS~17311--4924. According to Kozok's
data, the mean magnitudes are

$U=10^{m}.518$, $B=11^{m}.080$, $V=10^{m}.677$.

The mean brightness from the ASAS data for 2001-2009 was
$V=10^{m}.84\pm0.06$.

The $V$ brightness of IRAS~17311--4924 has slightly decreased
since the beginning of the past century. However, given the star's
rapid variability with a maximum amplitude up to 0$^{m}$.4 and the
large error of the magnitude from the CD catalogue, this assertion
cannot be considered quite justified.

{\bf IRAS~18023--3409.} For
CD--34$^{\circ}$12448=IRAS~18023--3409, we estimated its magnitude
at the epoch of the CD catalogue to be $V=10^{m}.8\pm0.8$. Despite
the significant uncertainty of this magnitude, it can be said with
complete confidence that the star at the beginning of the past
century was considerably brighter than at the present epoch.

Drilling et al. (1991) observed the star in 1972--1976 and
obtained the following magnitude estimates for it:

$U=11^{m}.70$, $B=12^{m}.01$, $V=11^{m}.55$.

According to the data from Tycho-2 (Hog et al. 2000), in 1989-1992
the star had $B=12^{m}.05\pm0.15$ and $V=12^{m}.08\pm0.26$.

According to the ASAS data, the star's mean magnitude in 2001-2009
was $V=12^{m}.41\pm0.10$, while the yearly mean magnitudes showed
a systematic fading.

For clarity, Fig. 6 shows the light curve for IRAS~18023--3409
from archival data. The modified $V$ magnitude from the CD
catalogue, the data from Drilling et al. (1991) and Hog et al.
(2000), and the yearly mean $V$ magnitudes of the star from the
ASAS data are plotted in the figure.

\begin{figure}
\centering
\includegraphics[scale=1.0]{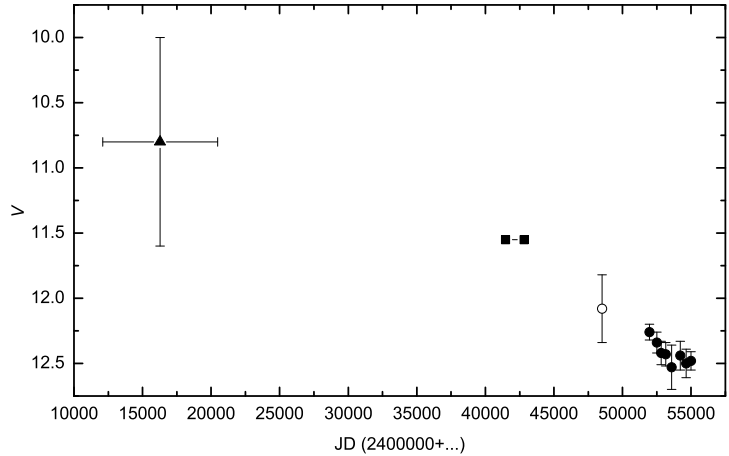}
\caption{Light curves of IRAS~18023--3409 from archival data. The
triangle in the figure designates the modified $V$ magnitude from
the CD catalogue; the data from Drilling et al. (1991) and Hog et
al. (2000) are indicated by the squares and the open circle,
respectively; the yearly mean $V$ magnitudes of the star from the
ASAS data are indicated by the filled circles.} \label{fig6}
\end{figure}

\section*{DISCUSSION}

According to the theory of stellar evolution, the duration of the
post-AGB phase depends on the initial mass of the star, the
history of its mass loss on the AGB, and the mass of the remnant
star. Bl\"{o}cker (1995) showed that for a certain set of these
parameters, there exists a possibility of very rapid evolution of
the star after the termination of large-scale mass loss on the AGB
before the beginning of the planetary nebula phase. For example,
for the model ($M_{ZAMS}$, $M_{c}$) = (5 $M_{\odot}$, 0.836
$M_{\odot}$), its lifetime in the post-AGB phase (6000 K $<
T_{eff} < $25~000 K) is $\sim$100 years. Therefore, the
possibility of detecting evolutionary changes in certain stars on
a time scale of several tens of years is not ruled out.

For our study, we selected hot post-AGB stars, early B supergiants
with IR excesses and emission lines in the spectrum, for which
Mello et al. (2012) obtained masses exceeding the mean masses of
planetary nebula nuclei.

Mello et al. (2012) determined the parameters of nine hot post-AGB
stars (with the stars from our sample being also among them) and
used the ($\log g - T_{eff}$) diagram to estimate the masses of
the stars and their progenitors using evolutionary tracks from
Sch\"{o}nberner (1983) and Bl\"{o}cker (1995). For
IRAS~17074--1845, the mass of the star was found to be $M_{c}=
0.63 M_{\odot}$. According to the estimates by Mello et al.
(2012), IRAS~17311--4924 and IRAS~18023--3409 have even higher
masses: $M_{c}=0.84M_{\odot}$.

An additional argument for a high mass of IRAS~18023--3409 is the
fact that the object exhibits an enhanced abundance of nitrogen
and helium and is the only star from the sample of Mello et al.
(2012) that satisfies the criteria for Peimbert's type I planetary
nebulae, namely (He/H)$>$0.14 or $\log$ (N/O)$>$0 (Peimbert and
Torres-Peimbert 1983). Intermediate mass (3.5 $M_{\odot}<M_{ZAMS}<
5.8 M_{\odot}$) stars are believed to be the progenitors of their
nuclei.

What observational manifestations of the rapid evolution of a star
in the post-AGB phase might be expected?

In the post-AGB phase, the optical brightness of the star will
decrease systematically at a constant bolometric luminosity with
increasing temperature starting from $T_{eff}\sim5700$ K. For
sufficiently hot stars that can already begin to ionize their
gaseous envelopes, the emission line intensities and the gas
ionization fraction are expected to increase.

{\bf IRAS~17074--1845.} The photometric observations of
IRAS~17074--1845 showed that the star faded in the $V$ band from
the epoch of the BD catalogue to the present time by more than 1
mag, corresponding to an increase in the star's temperature by
7000-8000 K under the assumption of evolution at a constant
bolometric luminosity. The rate of increase in temperature can
then be estimated to be $\sim$50 K yr$^{-1}$ and the star's
temperature could rise by $\sim$1000 K in 20 years of its
systematic spectroscopic observations. This value is too low to be
detectable from the available spectroscopic data. The difference
in the estimates of the spectral type for IRAS~17074--1845 from
B3I to B5I apparently reflects the difference between the criteria
for spectral classification. The emission line equivalent widths
from 1994 to 2012 showed no changes exceeding the measurement
errors.

{\bf IRAS~17311--4924.} The archival photometric data revealed a
slight, by no more than 0$^{m}.5$, fading of the star in the $V$
band from the epoch of the CD catalogue to the present time.

The estimates of the spectral type for IRAS~17311--4924 made by
Parthasarathy et al. (2000) based on the 1994 observations, B1IIe,
and those obtained by Carmona et al. (2010) based on the 2007
spectrum, B1Iae, agree well with the stellar parameters derived by
Mello et al. (2012), $T_{eff}=20~500\pm500$ K and $\log g$ = 2.35,
from the 2008 observations. Thus, the stellar parameters were
virtually constants in 14 years. However, the emission lines
belonging to the gaseous envelope show significant variations. For
example, according to our data, the equivalent widths of the [O~I]
and [N~II] lines increased from 1993 to 2012 by more than a factor
of 1.5. Mello et al. (2012) pointed to an increase in the
H$\alpha$ intensity in their spectrum compared to the data from
Sarkar et al. (2005). These changes may be related to the stellar
wind instability.

{\bf IRAS~18023--3409.} IRAS~18023--3409 showed a change in its
spectrum and brightness. The star's brightness declines
systematically: the star has faded in the $V$ band $\sim1^{m}$.5
since the beginning of the past century and continues to fade at
the present epoch.

According to the published data, the spectral type of the star
changed from B9I (Vijapurkar and Drilling 1993) to B2I
(Parthasarathy et al. 2000) in 16 years from 1978 to 1994,
suggesting a considerable, $\sim$10 000 K (!), increase in the
star's temperature. However, the star's effective temperature
$T_{eff}$ = 19 400 K estimated from the spectrum taken 14 years
later (Mello et al. 2012) corresponds to a spectral type B2-B3I
and this estimate does not confirm the further expected rise in
temperature. Unfortunately, the star's spectral type in 2012
cannot be estimated from our spectra.

If the change in spectral type from B9 to B2 is assumed to be real
and related to evolution, then the corresponding change in $V$
brightness must be equal to the difference of the bolometric
corrections $\Delta V=BC(\mathrm{B9})-BC(\mathrm{B2})=1^{m}.4$. In
reality, the brightness decline for IRAS~18023--3409 between the
two epochs of spectroscopic observations did not exceed $\Delta
V\sim 0^{m}.5$.

The data are insufficient to investigate the behavior of the
emission lines in the spectrum of IRAS~18023--3409 belonging to
the gaseous envelope. Based on a qualitative description of the
spectrum, we can only conclude that the H$\beta$ line in the
spectrum of IRAS~18023--3409 changes significantly with time: in
the spectrogram taken in 1994 and published in Vijapurkar et al.
(1998), H$\beta$ is an emission line, while it is an absorption
one in our 2012 spectrum, although the emission spectrum of the
envelope was expected to be enhanced with rising temperature of
the star. Mello et al. (2012) point out that H$\beta$ has a P~Cyg
profile. The variability of the H$\beta$ emission line is most
likely related to the wind activity of the star.

The source IRAS~18023--3409 showed the most significant changes in
brightness and spectrum among the stars from our sample. The
conclusion reached by Mello et al. (2012) about the star's high
mass ($M_{c}=0.82M_{\odot}$) based on their comparison of the
measured stellar parameters ($T_{eff}$ and $\log g$) with
theoretical evolutionary tracks from Bl\"{o}cker (1995) and the
enhanced abundance of nitrogen and helium in the stellar
atmosphere typical of the most massive planetary nebula nuclei
allows the revealed brightness changes in IRAS~18023--3409 to be
associated with the star's rapid evolution in the post-AGB phase.

\section*{CONCLUSIONS}

We obtained spectroscopy for three southern-hemisphere hot
post-AGB objects. We compared the new spectroscopic data with
those published previously. We analyzed the new photoelectric
observations of IRAS~17074--1845 and the archival photometric data
for all three sample stars.

The main conclusions of our study as as follows.

(1) Rapid photometric variability of IRAS~17074--1845,
IRAS~17311--4924, and IRAS~18023--3409 has been detected for the
first time. According to the ASAS data for 2001-2009, all three
objects exhibited brightness variations with an amplitude up to
$\Delta V\sim 0^{m}.3-0^{m}.4$ and a time scale of several days.
Such a pattern of variability has also been pointed out previously
for other hot post-AGB stars (Hrivnak et al. 2000; Arkhipova et
al. 2013b). It was hypothesized that an unsteady stellar wind was
mainly responsible for the brightness variations of hot
protoplanetary stars. Its origin can be associated with density
inhomogeneities in the outer layers of the star, a low-mass
supergiant that has passed the hydrogen and helium shell burning
stage on the AGB(Arkhipova et al. 2012).

(2) Our photoelectric $UBV$ observations of IRAS~17074--1845 in
2012-2013 confirmed the photometric variability of the star
revealed by the ASAS data. We found a correlation of the $B-V$ and
$U-B$ colors with the magnitude: an increase in $B-V$ and a
decrease in $U-B$ as the star brightens.

(3) Based on the archival data, we traced the photometric history
of the program objects. All three stars exhibited a fading on a
time scale longer than 100 years. IRAS~17074--1845 and
IRAS~18023--3409 showed the most prominent fading. Their $V$
brightness decreased by more than 1$^{m}$ compared to the epochs
of the BD and CD catalogues, respectively.

(4) For IRAS~17311--4924 and IRAS~18023--3409, we obtained data on
the change in their spectra over the last 20 years. For
IRAS~17311--4924, the equivalent widths of the [N~II] and [O~I]
emission lines increased noticeably. For IRAS~18023--3409, we
revealed a change in the spectral type and the H$\beta$ line.

Our hypothesis about rapid evolution of the objects presented here
is based on the detected brightness variations in the stars. The
secular fading in the $V$ band is consistent with an increase in
temperature at a constant luminosity in the post-AGB phase of
evolution. However, the spectrum variations can be associated both
with the evolutionary change in stellar temperature and with the
stellar wind instability. Therefore, spectroscopic observations on
a time scale longer than 20 years are needed to reach a more
definitive conclusion about the possible rapid evolution.

\bigskip

\textbf{ACKNOWLEDGMENTS}

This study was supported by the National Research Foundation of
the Republic of South Africa. We wish to thank the administration
of the South African Astronomical Observatory for the allocation
of observing time on the 1.9-m telescope. We used the 2MASS and
ASAS archival data.

\bigskip

\textbf{ REFERENCES}

\begin{enumerate}

\item V. P. Arkhipova, N. P. Ikonnikova, R. I. Noskova, et al.,
Astron. Lett. {\bf 25}, 25 (1999).

\item V. P. Arkhipova, N. P. Ikonnikova, and G. V. Komissarova,
Astron. Lett. {\bf 36}, 269 (2010).

\item V. P. Arkhipova, M. A. Burlak, V. F. Esipov, et al., Astron.
Lett. {\bf 38}, 157 (2012).

\item V. P. Arkhipova, N. P. Ikonnikova, A. Yu. Knyazev, et al.,
Astron. Lett. {\bf 39}, 201 (2013a).

\item V. P. Arkhipova, M. A. Burlak, V. F. Esipov, et al., Astron.
Lett. {\bf 39}, 619 (2013b).

\item C. A. Beichmann, G. Neugebauer, H. J. Habing, P. E. Clegg,
and T. J. Chester, {\it IRAS Point Source Catalogue}, JPL (1985).

\item T. Bl\"{o}cker, Astron. Astrophys. {\bf 299}, 755 (1995).

\item A. Carmona, M. E. van den Ancker, and M. Audard, Astron.
Astrophys. {\bf 517}, A67 (2010).

\item L. Cerrigone, J. L. Hora, G. Umana, and C. Trigilio,
Astrophys. J. {\bf 703}, 585 (2009).

\item Y.-S. Dong and J.-Y. Hu, Chin. Astron. {\bf 15}, 275 (1991).

\item J. S. Drilling, Astrophys. J. Suppl. Ser. {\bf 76}, 1033
(1991).

\item N. Epchtein, E. Deul, S. Derriere, et al., Astron.
Astrophys. {\bf 349}, 236 (1999).

\item P. Garc\'{i}a-Lario, M. Parthasarathy, D. de Martino, L.
Sanz Ferna. ndez de Co. rdoba, R. Monier, A. Manchado, and S. R.
Pottasch, Astron. Astrophys. {\bf 326}, 1103 (1997).

\item G. Gauba, M. Parthasarathy, Â. Kumar, R.K.S. Yadav and R.
Sagar, Astron. Astrophys. {\bf 404}, 305 (2003).

\item G. Gauba and M. Parthasarathy, Astron. Astrophys. {\bf 407},
1007 (2003).

\item G. Gauba and M. Parthasarathy, Astron. Astrophys. {\bf 417},
201 (2004).

\item G. Handler, R.H. Mendez, R. Medupe, et al., Astron.
Astrophys. {\bf 320}, 125 (1997).

\item K.G. Henize, Astrophys. J. Suppl. Ser. {\bf 30}, 491 (1976).

\item E. Hog, C. Fabricius, V.V.Makarov, Astron. Astrophys., {\bf
355}, L27 (2000).

\item B.J. Hrivnak, S. Kwok and  K.M. Volk, Astrophys. J. {\bf
346}, 265, (1989).

\item B.J. Hrivnak and W. Lu {\it {The Carbon Star Phenomenon,
Proceedings of the 177th Symposium of the International
Astronomical Union}, Vol. 177} ed. R. F. Wing (Dordrecht: Kluwer
Academic Publishers), p.293 (2000)

\item B.J. Hrivnak and W. Lu, R. E. Maupin, and B. D. Spitzbart
Astrophys. J., {\bf 709}, 1042, (2010).

\item L.L. Kiss, A. Derekas,  G.M. Szabo , T.R. Bedding and L.
Szabados, MNRAS {\bf 375}, 1338 (2007).

\item G. Klare, T. Neckel, Astron. and Astrophys., Suppl. Ser.,
{\bf 27}, 215 (1977).

\item A.Y. Kniazev, S.A. Pustilnik, E.K. Grebel, H. Lee, A.G.
Pramskij, Astrophys. J. Suppl. Ser. {\bf 153}, 429 (2004).

\item J. R. Kozok, Astron. Astrophys., Suppl. Ser. {\bf 61}, 387
(1985).

\item K. R. Lang, {\it Astrophysical Data: Planets and Stars}
(Springer) (1991).

\item V. M. Lyutyi, Soobshch. GAISh 172, 30 (1971).

\item D.R.C. Mello, S. Daflon, C.B. Pereira and I. Hubeny Astron.
Astrophys. {\bf 543}, A11 (2012).

\item M. Parthasarathy and S.R. Pottasch, Astron. Astrophys. {\bf
154}, L16 (1986).

\item M. Parthasarathy and S.R.Pottasch, Astron. Astrophys. {\bf
225}, 521 (1989).

\item M. Parthasarathy, ASPC {\bf 45}, 173 (1993).

\item M. Parthasarathy, P. Garc\'{i}a-Lario, S.R.Pottasch et al.,
Astron. Astrophys. {\bf 267}, L19 (1993).

\item M. Parthasarathy, J. Vijapurkar, J.S. Drilling, Astron.
Astrophys. Suppl. Ser. {\bf 145}, 269 (2000).

\item M. Peimbert and S. Torres-Peimbert, in  \textit{Proceedings
of the IAU Symposium ¹ 103 on Planetary Nebulae}, Ed. by
D.R.Flower (D. Reidel, Dordrecht, 1983), p. 233.

\item G. Pojmanski, Acta Astronomica {\bf 52}, 397 (2002).

\item A. Preite-Martinez, Astron. Astrophys. Suppl. Ser. {\bf 76},
317 (1988).

\item G. Sarkar, M. Parthasarathy And B.E. Reddy, Astron.
Astrophys., {\bf 431}, 1007 (2005).

\item D.J. Schlegel, D.P. Finkbeiner, and M. Davis, Astrophys. J.
{\bf 500}, 525 (1998).

\item D. Sch\"{o}nberner, Astrophys. J. {\bf 272}, 708 (1983).

\item D. Sch\"{o}nberner, \textit{Late Stages of Stellar
Evolution}, Ed. by S. Kwok and S. R. Pottasch (Reidel, Dordrecht,
1987), p. 341.

\item M. F. Skrutskie, R. M. Cutri, R. Stiening et al., Astron.
J., {\bf 131}, 1163 (2006).

\item C. B. Stephenson and N. Sanduleak, \textit{Publication of
the Warner and Swasey Observatory} (Case Western Reserve Univ.,
Cleveland, Ohio, 1971).

\item V. L. Strai\v{z}his, \textit{Metal-Deficient Stars}
(Mokslas,Vilnyus, 1982) [in Russian].

\item O. Su\'{a}rez, P. Garc\'{i}a-Lario, A. Manchado, M.
Manteiga, A. Ulla  and S.R. Pottasch, Astron. Astrophys. {\bf
458}, 173 (2006).

\item R. Szczerba, N. Siodmiak, G. Stasinska  and J. Borkowski,
Astron. Astrophys. {\bf 469}, 799 (2007).

\item W. E. C. J. van der Veen, H. J. Habing, and T. R. Geballe,
Astron. Astrophys. {\bf 226}, 108 (1989).

\item J. Vijapurkar and J. S. Drilling, Astrophys. J. Suppl. Ser.
{\bf 89}, 293 (1993).

\item J. Vijapurkar, M. Parthasarathy, and J. S. Drilling, Bull.
Astron. Soc. India {\bf 26}, 497 (1998).

\item  P. A. Wehinger and B. Hidajat, Astron. J. {\bf 78}, 401
(1973).

\item H. van Winckel, T. L. Evans, M. Briquet, P. De Cat, P.
Degroote, W. De Meester, J. De Ridder, P. Deroo, et al., Astron.
Astrophys. {\bf 505}, 1221 (2009).

\item H. van Winckel, B. J. Hrivnak, N. Gorlova, C. Gielen, and W.
Lu, Astron. Astrophys. {\bf 542}, 53 (2012).

\end{enumerate}

\newpage

\begin{table*}
\centering

{\bf Erratum: {\bf Variability and Possible Rapid Evolution of the
Hot Post-AGB Stars Hen 3-1347, Hen 3-1428, and LSS 4634}
[Astronomy Letters, 2014, {\bf 40}, 485 (2014)].}

\bigskip

{\it V. P. Arkhipova, M.A.Burlak, V. F. Esipov, N. P. Ikonnikova,
A. Yu. Kniazev,\\ G. V. Komissarova, and A. Tekola}

\end{table*}

\bigskip

In this note, we add two references mentioned in the text, but
missed in the list of references.

\bigskip

REFERENCES

\begin{enumerate}

\item M. Parthasarathy and S.R.Pottasch, Astron. Astrophys. {\bf
225}, 521 (1989).

\item M. Parthasarathy, ASPC {\bf 45}, 173 (1993).

\end{enumerate}
{\bf Acknowledgments}

We thank M. Parthasarathy for pointing out missing references.

\end{document}